\documentclass[12pt,psfig]{article}
\usepackage{amssymb}
\usepackage{amsmath,amsthm}
\usepackage{amsfonts}
\usepackage{graphicx}
\usepackage{appendix}
\usepackage{braket}
\usepackage{graphics}
\usepackage{indentfirst}
\numberwithin{equation}{section}

\usepackage{hyperref}
\usepackage{accents}

\newcommand{\ubar}[1]{\underaccent{\bar}{#1}}

\begin{document}

\

\begin{center}
{\huge \textbf{LREE of a Dressed D$p$-brane 
with Transverse Motion in the Partially Compact Spacetime
}}\\
\vspace*{.95cm}
{\Large  \textsc{Hamidreza Daniali}\footnote{\textsc{Email:} \ 
\textit{hrdl@aut.ac.ir}} \&  \textsc{Davoud Kamani}
\footnote{\textsc{Email:} \  
\textit{kamani@aut.ac.ir} (Corresponding author)}}\\

\vspace*{.5cm}
\it{Department of Physics, Amirkabir University of
Technology (Tehran Polytechnic) \\
P.O.Box: 15875-4413, Tehran, Iran \\}
\end{center}
\vspace*{1.25cm}
\begin{abstract}
  In the context of the type IIA/IIB superstring theories, 
  we derive the left-right entanglement entropy (LREE) of a 
  BPS D$p$-brane with transverse motion in the presence of a 
  $U(1)$ gauge potential and the Kalb-Ramond field in the 
  partially compact spacetime
  $\mathbf{T}^n \otimes \mathbb R^{1,9-n}$. 
  At first we employ the replica trick to compute 
  the R\'{e}nyi entropy and then we obtain the entanglement 
  entropy. We examine the results for the special case,
  i.e. for the D6-brane. Besides, we investigate the 
  thermodynamical entropy, associated with the LREE. This  
  demonstrates that the LREE is precisely equivalent 
  to its thermodynamic counterpart. 
   
\end{abstract}
\begin{flushleft}
\textsc{PACS numbers:} 11.25.-w; 11.25.Uv\\
\vspace*{.25cm}
\textsc{Keywords:} Boundary state; Interaction amplitude;
Compact spacetime; R\'{e}nyi entropy; LREE;
Thermodynamical entropy. 
\end{flushleft}

\newpage

\section{Introduction}
\label{100}

Entanglement is indeed a pivotal 
characteristic of quantum mechanics. 
It reflects the non-local correlations 
between subsystems of a composite 
quantum system, such that the quantum 
state of each subsystem cannot be 
characterized independently of the others. 
For composite quantum systems in pure 
states, a useful tool for 
describing the entanglement is the entanglement entropy. 
It specifies how much information is shared between 
different parts of a quantum system.

In the context of the AdS/CFT correspondence, 
the Ryu-Takayanagi formula offers a 
holographic representation of the entanglement 
entropy \cite{1}. This suggests that 
the entanglement of the quantum fields in the 
boundary theory possesses a direct geometric 
interpretation in the bulk spacetime. 
It is related to the areas of 
minimal surfaces in the anti-de Sitter space. 
This insight is naturally extended 
to the black holes thermodynamics and 
the Bekenstein-Hawking entropy \cite{2}. 
Beyond these, the entanglement entropy 
has found significant utility in the 
condensed matter systems. It has been applied in 
systems which exhibit topological 
order or critical behavior \cite{3}. 
In the many-body quantum systems, 
it distinguishes between the different 
phases by characterizing the scaling 
of the entanglement with the system size \cite{4}.

In the conventional procedures, the 
physical separation of the entangled subsystems 
gives a corresponding separation in the Hilbert space. 
Nevertheless, in this paper we investigate 
an alternative method in which the 
partitioning of the subsystems is 
exclusively defined within the Hilbert space. 
It regards the left- and right-moving modes of a closed 
superstring as distinct subspaces. 
The entanglement entropy that 
quantifies the correlation between 
the left- and right-moving modes 
is known as the left-right 
entanglement entropy (LREE) 
\cite{5}-\cite{9}. This conceptual framework 
introduces a rigorous entanglement in the 
string theory. It emphasizes 
the internal degrees of freedom 
rather than the spatial separation.

On the other hand, we have the  
D-branes in the string theory. 
The dynamics of the D-branes are 
associated with key domains 
such as the AdS/CFT correspondence, 
string phenomenology, and black hole 
physics \cite{10} , \cite{11}. 
Boundary states have been widely 
used for the description of the D-branes 
systems \cite{12}-\cite{21}. 
We intend to investigate the LREE 
of a D$p$-brane by employing its corresponding 
boundary state.

Note that the LREE was initially investigated for
a one-dimensional boundary state in a 2D 
CFT with the Dirichlet/Neumann  boundary 
condition \cite{5}. Then, it was extended  
to a  bare-stationary D$p$-brane \cite{6}. 
Besides, some other configurations, including the bosonic 
dressed-dynamical D$p$-brane 
\cite{22} and  its non-BPS counterpart \cite{23},  and 
also the time-dependent plane wave in the  
context of GS superstring theory 
\cite{8}, have been investigated. 
Only a few of them explicitly incorporate  the 
R-R sector in their configurations. 
In addition, they have not considered
the effects of the spacetime compactification.

Our motivation comes from the effects
of the R-R sector, accompanied by  
the spacetime compactification, on LREE.  
Precisely, in the context of 
type IIA/IIB superstring theories, we shall 
use the boundary state formalism to 
compute the LREE of a BPS stable 
D$p$-brane. Our brane has a motion  
along a transverse direction to itself,
and it has been dressed with a $U(1)$ gauge potential 
and a Kalb-Ramond field, in the partially compact spacetime  
$\mathbf{T}^n\otimes \mathbb{R}^{1, 9-n}$. 

This paper is organized as follows. 
In Sec. \ref{200}, we review the density 
matrix formulation in the context of 
a D$p$-brane via its corresponding boundary state. 
In Sec. \ref{300}, we compute the 
interaction amplitude between two identical 
parallel BPS D$p$-branes within our 
setup to derive the partition function. 
In Sec. \ref{400}, the LREE is explicitly derived  
by applying an appropriate limit of the R\'{e}nyi entropy. 
Besides, we shall examine the 
result for the critical dimension $p=6$. 
In addition, the equivalence between 
thermal entropy and LREE will be studied. Sec. 
\ref{500} is devoted to the conclusion. 
The computations of the boundary state, 
associated with the D$p$-branes of our setup, 
will be provided in the Appendix \ref{app}.

\section{The density matrix formulation for a D$p$-brane}
\label{200}
 
In this section, we apply the 
conventional notations which are usually used for 
the composite quantum systems. 
Assume that the composite system only includes  
the subsystems $\mathcal{A}$ and $\mathcal{B}$.
The density operator is represented as 
$\rho = |\Psi\rangle\langle \Psi|$, 
where $\Psi$ corresponds to the 
pure state of the composite system. Besides, 
the reduced density operator is defined as 
$\rho_{\mathcal{A}} \equiv \text{Tr}_{\mathcal{B}} \rho$, 
where $\text{Tr}_{\mathcal{B}}$ 
indicates the partial trace over the 
subsystem $\mathcal{B}$.

In a composite quantum system, the 
quantum state of each subsystem clearly is 
interdependent on the states of the other subsystems. 
The entanglement and R\'{e}nyi entropies are  
effective tools for calculating the entanglement
between the subsystems. The von Neumann formula, ${\bf S}
=-{\rm Tr}(\rho_{\mathcal A} \ln \rho_{\mathcal A})$, 
can be used to calculate the former 
quantity, while the latter one is derived 
from ${ \bf S}_n|^{n>0}_{n \ne 1}
=(1-n)^{-1} \ln {\rm Tr}\rho_{\rm A}^n$. 
In the limit $n \rightarrow1$, the R\'{e}nyi 
entropy approaches the entanglement entropy.

To translate these in the context of the 
closed superstring theory, one can always  
specify the associated Hilbert 
space as a tensor product of two subspaces 
$\mathbb{H}=\mathbb{H}_{\rm L}\otimes \mathbb{H}_{\rm R}$.
This elegant structure originates from the 
decomposition of the oscillatory degrees of 
freedom of closed superstring 
into the left- and right-moving modes. 
These modes serve as the basis states 
for the subsystems with the labels  
``L'' and ``R'', respectively.  
To obtain the physical subspace of this Hilbert space, 
it is necessary to impose the Virasoro constraints. 
A generic state in the Hilbert 
space of the closed superstring theory can be 
explicitly expressed as 
$|\psi\rangle = |\psi\rangle_{\rm L} 
\otimes |\psi\rangle_{\rm R}$, 
where the individual states for the 
left- and right-moving parts are given by
\begin{eqnarray}
|\psi\rangle_{\rm L} &=& \prod_{k=1}^\infty \prod_{t} 
\frac{1}{\sqrt{n_k !}} 
\left(\frac{\alpha^{\mu_k}_{-k}}{\sqrt{k}}\right)^{n_k} 
\left(\psi^{\mu_t}_{-t}\right)^{n_t}|0\rangle,  \\
|\psi\rangle_{\rm R} &=& \prod_{k=1}^\infty \prod_{t} 
\frac{1}{\sqrt{m_k !}} 
\left(\frac{\tilde{\alpha}^{\nu_k}_{-k}}
{\sqrt{k}}\right)^{m_k} 
\left(\tilde{\psi}^{\nu_t}_{-t}\right)^{m_t}|0\rangle, 
\end{eqnarray}
where, in the R-R (NS-NS) sector, the mode indices 
``$t$''s are positive integers (half integers) numbers. 
Since the oscillators $\psi_{-t}^\mu$ 
and $\tilde{\psi}_{-t}^\nu$ are Grassmann-valued, 
the occupation numbers $m_t$ and 
$n_t$ are restricted to the set $\{0, 1\}$. 
The integer numbers 
$\{n_t , n_k , m_t , m_k | k \in \mathbb{N}\}$ 
are mutually independent, except under the constraint 
$\sum_{k=1}^\infty k n_k + 
\sum_{t} t n_t = \sum_{k=1}^\infty k m_k + \sum_{t} t m_t$. 
The LHS and RHS of this condition prominently represent 
the total mode numbers in the 
states $|\psi\rangle_{\rm L}$ and 
$|\psi\rangle_{\rm R}$, respectively. 
Thus, the Virasoro constraints 
clearly impose the equality of the 
total mode numbers between the 
left- and right-moving parts. 
This condition enforces a mild 
correlation between the left- and right-moving modes, 
otherwise they remain decoupled. 
Consequently, the physical Hilbert 
space preserves its factorized form, 
consistent with the independence of 
the left- and right-moving sectors.

The boundary state, which represents 
a coherent state within the framework of 
closed string theory, can be 
decomposed into its left- and right-moving modes through 
the Schmidt decomposition \cite{24}. 
This decomposition effectively reveals 
an entanglement structure in the 
boundary state \cite{5}, \cite{6}. The expansion of the 
exponential factors in Eqs. \eqref{a7}, \eqref{a16} 
and \eqref{a17} leads to a series 
that explicitly specifies the entangled 
nature between the left- and right-moving 
parts of the Hilbert space.

The density operator, associated with the 
boundary state $|B\rangle$, may be expressed as 
$\rho = |B\rangle \langle B|$. However, 
the norm $\langle B|B\rangle$ obviously is divergent 
which gives ${\rm Tr}\,\rho \ne 1$. 
Hence, we introduce a regularized state of the form 
$|\mathbb{B}\rangle = \mathcal{N}_B^{-1/2 } 
e^{-\epsilon H} |B\rangle $, 
where $\epsilon$ represents a finite 
correlation length, and the normalization factor 
$\mathcal{N}_B$ is determined by enforcing 
the probability conservation condition. 
Besides, $H$ is the total Hamiltonian of 
the propagating closed superstring. 
The corresponding density operator to this state is  
$\rho = \mathcal{N}^{-1}_B \Big(e^{-\epsilon H} 
|B\rangle \langle B| e^{-\epsilon H}\Big)$.
Upon taking the trace of the density 
operator over the closed superstring 
states and applying the condition 
${\rm Tr}\,\rho = 1$, the normalization 
factor finds the value  
$\mathcal{N}_B = \langle B|e^{-2\epsilon H}| B\rangle$. 
Therefore, one can conveniently read the 
normalization factor from the interaction of two identical 
D$p$-branes at zero distance, 
in which the propagating closed superstring 
moves between them for the time $t=2\epsilon$.

\section{The interaction amplitude and partition function}
\label{300}

In order to determine the interaction 
amplitude of a D$p$-D$p$ system, we compute 
the tree-level diagram of a 
closed string that propagates between the 
branes. At first, we consider the general case, 
in which each D$p$-brane possesses its own fields 
and transverse dynamics. Subsequently, 
we shall obtain the partition function. 

Now we define the following notations. 
$\alpha, \beta \in \{0,1,\cdots, p\}$ 
represent the worldvolume directions of the D$p$-branes. 
The subset $\{a, b \} = \{ \alpha, \beta\} 
- \{ 0 \}$ belongs to the spatial directions
of the brane. The index $i_{\mathcal V}$ is used to 
indicate the boost direction. In addition, the indices 
$i,j \in \{p+1,\cdots, 9\}$ show the orthogonal directions
to the worldvolume of the D$p$-brane, and 
the parameters $y^i$ specify its position. 
Besides, the overbarred indices are 
used to denote the compact directions, 
whereas the underbarred indices exhibit the 
non-compact ones.

The amplitude can be determined by 
computing the overlap of the total 
boundary states via the propagator 
$\mathcal D$ of the exchanged closed superstring
\begin{eqnarray}
{\bf A} =\ _{\rm tot}\langle B_1| 
\mathcal D |B_2\rangle_{\rm tot}\;, 
\qquad \mathcal D = 2\alpha' 
\int_0^{\infty} {\rm d}t \ e^{-tH}, 
\label{1}
\end{eqnarray}
The total boundary state 
$|B\rangle_{\rm tot}$ is formed by 
the direct product of the bosonic portion, 
given by Eqs. \eqref{a7} and 
\eqref{a11}, and the conformal $(b,c)$-ghosts, 
along with the GSO-projected versions of 
Eqs. \eqref{a16}-\eqref{a21}, and 
their corresponding $(\beta, \gamma)$-superghosts. 
With the provided details in the Appendix 
\ref{app} and extensive calculations, 
the interaction amplitude of the
two dressed D$p$-branes with the 
different transverse velocities in the 
$\bf{T}^n \otimes \mathbb R^{1,9-n}$ 
spacetime takes the feature
\begin{eqnarray}
\mathbf{A} &=& \frac{\alpha' V_a 
T_p^2 \sqrt{\det(\mathcal M_1 
\mathcal M_2)}}{8 (2\pi)^{d_{\ubar{i}}} 
|\mathcal V_1 - \mathcal V_2|} 
\int_0^\infty {\rm d}t \left(\frac{\pi}
{\alpha' t}\right)^{d_{\ubar{i}}} 
\exp\left(-\frac{1}{4\alpha' t} 
\sum_{\ubar{i}} \big( y_1^{\ubar{i}} 
- y_2^{\ubar{i}} \big)^2 \right)
\nonumber\\
&\times&  \left\{ \left(\frac{e^t}{4}
\right)^2 \prod_{m=1}^\infty \left[
\left(\frac{1- e^{-4mt}}{1
+e^{-2(2m-1) t}}\right)^{p-6} \frac{\det(\mathbb{I}
+ \mathcal Q_1^T \mathcal Q_2 
e^{-2(2m-1)t})}{\det(\mathbb{I}
- \mathcal Q_1^T \mathcal Q_2 e^{-4mt})}\right. \right] 
\nonumber\\
&-& \left. \left(\frac{e^t}{4}\right)^2 \prod_{m=1}^\infty 
\left[\left(\frac{1- e^{-4mt}}{1-e^{-2(2m-1) t}}\right)^{p-6} 
\frac{\det(\mathbb{I}- \mathcal Q_1^T 
\mathcal Q_2 e^{-2(2m-1)t})}{\det(\mathbb{I}
- \mathcal Q_1^T \mathcal Q_2 e^{-4mt})}\right.\right]
\nonumber\\
&-& \left. \kappa \prod_{m=1}^\infty \left[
\left(\frac{1- e^{-4mt}}{1+e^{-4m t}}
\right)^{p-6} \frac{\det(\mathbb{I}
+ \mathcal Q_1^T \mathcal Q_2 
e^{-4mt})}{\det(\mathbb{I}
- \mathcal Q_1^T \mathcal Q_2 
e^{-4mt})}\right] - \kappa' \right\}
\nonumber\\
&\times& \Big( 1 + \sum_c 
\Big\{\exp\left[-\frac{t}{\alpha'} 
\mathcal{L}_c^{\bar a} \mathcal{L}_c^{\bar b} 
(\delta_{\bar a \bar b}
+ \mathfrak J^+_{\bar a} \mathfrak{J}^-_{\bar b} 
+ \mathcal F^{a}_{1\ \bar a} 
\mathcal F_{2 \ a \bar{b}}\right] 
\nonumber\\
&\times& \exp{\left[\frac{i}{\alpha'} 
\mathcal{L}_c^{\bar{a}} 
\Big(\mathfrak{P}_{\bar{a}} y_2^{i_{\mathcal V}} 
- \mathfrak{T} _{\bar{a}} 
y_1^{i_{\mathcal V}}\Big)\right]} \Big\} \Big)
\prod_{\bar i}\Theta_3 \left(\frac{y_1^{\bar{i}} 
- y_2^{\bar{i}}}{2\pi R_{\bar i}} 
\Big\lvert \frac{i\alpha' t}{\pi R^2_{\bar i}} 
\right), 
\label{3.2}
\end{eqnarray}
where $V_a$ is the common volume of the branes, and 
$\mathcal F_{\alpha\beta} 
= F_{\alpha \beta} - B_{\alpha\beta}$ 
is the total field strength. Besides, we have defined 
\begin{eqnarray}
\mathfrak{P}_{\bar{a}}&\equiv& \frac{1}{\mathcal V_2 
- \mathcal V_1} \Big[\gamma_2^2 
(1+\mathcal V_1 \mathcal V_2) 
\mathcal F_{2 \ \bar{a}}^0 
- \gamma_1^2 (1+ \mathcal V^2_1) 
\mathcal F_{1 \ \bar{a}}^0\Big],
\label{3.3}\\
\mathfrak{J}^\pm_{\bar a} &\equiv &
\frac{1}{|\mathcal V_1 - \mathcal V_2|} 
\Big[\gamma_2^2 (1 \pm \mathcal V_1 )(1 + \mathcal V_2^2)  
\mathcal F_{2 \ \bar{a}}^0  
- \gamma_1^2 (1 \pm \mathcal V_2 )(1 + \mathcal V_1^2)  
\mathcal F_{1 \ \bar{a}}^0\Big], \label{3.4}
\end{eqnarray}
and $\mathfrak{T} = \mathfrak P|_{\mathcal 
V_1 \leftrightarrow \mathcal V_2}$.

Let us clarify the amplitude \eqref{3.2}. The first 
and the last two lines arise from the zero-mode 
contribution to the boundary states. 
The exponential factor in the first 
line prominently shows the damping nature  
of the interaction, which depends on  
the squared distance between the 
branes. The subsequent two 
lines correspond to the oscillatory parts of the 
GSO-projected boundary states within the NS-NS sector. 
The fourth line belongs to the R-R sector. 
The parameters $(\kappa, \kappa')$, 
originate from the zero-modes 
of the R-R sector. These parameters are defined as follows
\begin{eqnarray}
\mathbf{A}^{0 \ \psi}_{R} (+,-) 
&=& \mathbf{A}^{0 \ \psi}_{R} 
(-,+) \equiv 2 \kappa, \\
\mathbf{A}^{0 \ \psi}_{R} (+,+) 
&=& \mathbf{A}^{0 \ \psi}_{R} 
(-,-) \equiv 2 \kappa'.
\end{eqnarray}
Thus, $\kappa$ and $\kappa'$ explicitly take the forms
\begin{eqnarray}
\kappa &=& \frac{1}{2} \mathrm{Tr}
\left\{C \mathbb{G}_{2} C^{-1} 
\mathbb{G}_{1}^{T} \left[-1 
+ \mathcal{V}_{1} \mathcal{V}_{2} 
+ (\mathcal{V}_{1} - \mathcal{V}_{2})
(\Gamma^{i_{\mathcal{V}}} 
\Gamma^{0})^{T}\right]\right\}, 
\label{3.7}\\
\kappa' &=& \mathrm{Tr}\left\{ C \mathbb{G}_{2} C^{-1} 
\mathbb{G}_{1}^{T} \left[-1 
+ \mathcal{V}_{1} \mathcal{V}_{2} 
+ (\mathcal{V}_{1} - \mathcal{V}_{2})
(\Gamma^{i_{\mathcal{V}}} 
\Gamma^{0})^{T}\right]\Gamma^{11}\right\}.
\label{3.8}
\end{eqnarray}
We should note that the four terms in Eq. \eqref{3.2} 
arise from the NS-NS, 
NS-NS$(-1)^{F}$, R-R, and R-R$(-1)^{F}$ 
sectors, respectively. The 
parameter $\kappa'$, for some brane 
configurations such as the stationary 
branes, usually vanishes. 
In other words, this quantity possesses a 
nonzero value only in specific 
configurations, such as in our case.

We employed the BPS branes.
The GSO projection removed the tachyon state from 
the NS-NS sector, hence we obviously acquired the stable
branes. Besides, due to the supersymmetry, 
the possible signs of $(\kappa, \kappa')$ in 
Eq. \eqref{3.2} correspond to 
interactions between brane-brane and 
antibrane-antibrane systems.

The last two lines of the amplitude 
\eqref{3.2} entirely reflect the 
effects of the compactification. 
By quenching the compactification effects, 
the interaction amplitude of two 
D$p$-branes in the non-compact spacetime is given by
\begin{eqnarray}
\mathbf{A}^{\rm non-compact} &=& \frac{\alpha' 
V_a T_p^2 \sqrt{\det(\mathcal{M}_1 
\mathcal{M}_2)}}{8 (2\pi)^{d_i} 
|\mathcal{V}_1 - \mathcal{V}_2|} 
\int_0^\infty {\rm d}t \left(\frac{\pi}{\alpha' t}
\right)^{d_i} \exp\left(-\frac{1}{4\alpha' t} 
\sum_i \left( y_1^i - y_2^i \right)^2 \right)
\nonumber\\
&\times&  \left\{\left(\frac{e^t}{4}
\right)^2 \prod_{m=1}^\infty 
\left(\frac{1- e^{-4mt}}{1+e^{-2(2m-1) t}}
\right)^{p-6} \frac{\det(\mathbb{I}
+ \mathcal{Q}_1^T \mathcal{Q}_2 
e^{-2(2m-1)t})}{\det(\mathbb{I}
- \mathcal{Q}_1^T \mathcal{Q}_2 e^{-4mt})}\right.  
\nonumber\\
&-& \left. \left(\frac{e^t}{4}\right)^2 \prod_{m=1}^\infty 
\left(\frac{1- e^{-4mt}}{1-e^{-2(2m-1) t}}\right)^{p-6} 
\frac{\det(\mathbb{I}- \mathcal{Q}_1^T \mathcal{Q}_2 
e^{-2(2m-1)t})}{\det(\mathbb{I}- \mathcal{Q}_1^T 
\mathcal{Q}_2 e^{-4mt})}\right.\nonumber\\
&-& \left. \kappa \prod_{m=1}^\infty 
\left(\frac{1- e^{-4mt}}{1+e^{-4m t}}\right)^{p-6} 
\frac{\det(\mathbb{I}+ \mathcal{Q}_1^T 
\mathcal{Q}_2 e^{-4mt})}{\det(\mathbb{I}
- \mathcal{Q}_1^T \mathcal{Q}_2 
e^{-4mt})} - \kappa' \right\},
\end{eqnarray}
which is in agreement with the conventional 
results in the literature.

To derive the partition function, 
it is necessary to fix the integral 
parameter as $ t \equiv 2\epsilon$. 
This implies that the interacting D$p$-branes 
are very close to each other. 
In fact, the calculation 
of the partition function elucidates 
that a D$p$-brane interacts with itself. 
Precisely, a closed superstring 
is emitted by the D$p$-brane, propagates 
for the time  $t = 2\epsilon $, 
and then is re-absorbed by the same 
D$p$-brane. This configuration can lead 
to a divergence. As we saw, 
the term $|\mathcal{V}_1 - \mathcal{V}_2|$ 
was appeared in the denominator. This 
is due to the identity 
$\delta(ax) = \delta(x)/|a|$. 
Hence, the partition function 
cannot be simply obtained by setting 
$\mathcal{V}_1 = \mathcal{V}_2 $. 
Therefore, it is essential to re-evaluate 
the interaction of the zero-mode 
component of the bosonic part of 
the boundary state in such conditions. 
Since the D$p$-branes are 
identical and are located at the same position, 
the indices 1 and 2 will be omitted, and 
the $y$-dependence will be removed. 
Consequently, the partition function takes the 
following feature 
\begin{eqnarray}
\mathcal{N}_B &=& \frac{\alpha' V_a 
T_p^2 \ |\det\mathcal{M}|}{8 
(2\pi)^{d_{\ubar{i}}} } \left(
\frac{\pi}{2\alpha' \epsilon }\right)^{d_{\ubar{i}}}
\left\{ \frac{1}{16q} \prod_{m=1}^\infty\left[ 
\left(\frac{1- q^{2m}}{1+q^{2m-1}}\right)^{p-6} 
\frac{\det(\mathbb{I}+ \mathcal{Q}^T 
\mathcal{Q} q^{2m-1})}{\det(\mathbb{I}
- \mathcal{Q}^T \mathcal{Q} q^{2m})}\right.\right]  
\nonumber\\
&-& \left. \frac{1}{16q}\prod_{m=1}^\infty \left[
\left(\frac{1- q^{2m}}{1-q^{2m-1}}\right)^{p-6} 
\frac{\det(\mathbb{I}- \mathcal{Q}^T 
\mathcal{Q} q^{2m-1})}{\det(\mathbb{I}
- \mathcal{Q}^T \mathcal{Q} q^{2m})}\right.\right]
\nonumber\\
&-& \left. \tilde{\kappa} \prod_{m=1}^\infty \left[
\left(\frac{1- q^{2m}}{1+q^{2m}}\right)^{p-6} 
\frac{\det(\mathbb{I}+ \mathcal{Q}^T 
\mathcal{Q} q^{2m})}{\det(\mathbb{I}
- \mathcal{Q}^T \mathcal{Q} q^{2m})} \right]
- \tilde{\kappa}' \right\}
\nonumber\\
&\times& \left( 1 + \sum_c \left\{ 
\exp\left[-\frac{2\epsilon }{\alpha'} 
\mathcal{L}_c^{\bar{a}} \mathcal{L}_c^{
\bar{b}} (\delta_{\bar{a} \bar{b}} 
+ \mathcal{F}^{a}_{\ \bar{a}} 
\mathcal{F}_{ \ a \bar{b})}\right] \right\} \right)  
\prod_{\bar{i}}\Theta_3 \left( 0 \ \Big\lvert \ 
\frac{2i\alpha' \epsilon}{\pi R^2_{\bar{i}}} \right), 
\label{3.10}
\end{eqnarray}
where the modular parameter is defined 
as $ q \equiv e^{-4\epsilon}$, 
and $ (\tilde{\kappa}, \tilde{\kappa}')$ are provided by 
Eqs. \eqref{3.7} and \eqref{3.8} in which 
$ \mathcal{F}_1 = \mathcal{F}_2 \equiv \mathcal{F}$ 
and $ \mathcal{V}_1 = \mathcal{V}_2 
\equiv \mathcal{V}$ should be applied. 
Using the Dedekind functions,
Eq. \eqref{3.10} can be expressed in 
a more elegant form
\begin{eqnarray}
\mathcal{N}_B &=& \frac{\alpha' V_a T_p^2 \ 
|\det\mathcal{M}|}{8 (2\pi)^{d_{\ubar{i}}} } 
\left(\frac{\pi}{2\alpha' \epsilon }\right)^{d_{\ubar{i}}}
\left\{ \frac{1}{16} q^{-(p-6)/8}\left[ 
\left(\frac{f_1(q)}{f_3(q)}\right)^{p-6} 
\prod_{m=1}^\infty \frac{\det(\mathbb{I}
+ \mathcal{Q}^T \mathcal{Q} q^{2m-1})}{\det(\mathbb{I}
- \mathcal{Q}^T \mathcal{Q} q^{2m})}\right. \right. 
\nonumber\\
&-& \left. \left. \left(\frac{f_1(q)}{f_4(q)}\right)^{p-6} 
\prod_{m=1}^\infty\frac{\det(\mathbb{I}
- \mathcal{Q}^T \mathcal{Q} 
q^{2m-1})}{\det(\mathbb{I}- 
\mathcal{Q}^T \mathcal{Q} q^{2m})}\right] 
- \tilde{\kappa} \left(\frac{f_1(q)}{f_2(q)}\right)^{p-6} 
\prod_{m=1}^\infty\frac{\det(\mathbb{I}
+ \mathcal{Q}^T \mathcal{Q} q^{2m})}{\det(\mathbb{I}
- \mathcal{Q}^T \mathcal{Q} q^{2m})}\right. 
\nonumber \\ 
&-& \left. \tilde{\kappa}' \right. \Bigg\} \left( 1 
+ \sum_c \left\{ \exp\left[-\frac{2\epsilon }{\alpha'} 
\mathcal{L}_c^{\bar{a}} \mathcal{L}_c^{\bar{b}} 
(\delta_{\bar{a} \bar{b}} + \mathcal{F}^{a}_{\ \bar{a}} 
\mathcal{F}_{ \ a \bar{b}})\right] \right\} \right)  
\prod_{\bar{i}}\Theta_3 \left( 0 \ \Big\lvert \ 
\frac{2i\alpha' \epsilon}{\pi R^2_{\bar{i}}} \right). 
\label{11}
\end{eqnarray}

\section{The associated LREE}
\label{400}

The first step for computing the LREE involves  
evaluation of the R\'{e}nyi entropy. 
Thus, it is necessary to 
determine ${\rm Tr}\,\rho_{{L}}^n$. 
Employing the replica trick,  
for the real values of ``$n$'' we receive  
${\rm Tr}\,\rho_{{L}}^n \sim \mathcal 
Z(2n\epsilon)/ \mathcal{N}_B $, 
in which $Z(2n\epsilon)$ is called 
\textit{replicated partition function}.
As stated in the Ref. \cite{5}, many 
approaches may be potentially exerted to 
calculate ${\rm Tr}\,\rho_{{L}}^n$ by 
summing over the spin structures 
and momentum. Nevertheless, our analysis 
is limited to the particular scenario that 
preserves the open/closed string duality, i.e., the 
unreplicated normalization constant, 
the correlated momentum and the correlated spin structure. 

By considering $\epsilon \rightarrow 0$, 
which explicitly indicates a single D$p$-brane, 
the modular parameter approaches
unity. This induces a trivial divergence, 
as previously mentioned in Sec. 
\ref{200}. In fact, when one considers 
an infinitesimal exponent of the modular 
parameter, the open string 
amplitude becomes more appropriate. 
Consequently, by employing the 
open/closed duality via the Jacobi transformation 
$q \rightarrow \hat{q}\equiv 
e^{-1/4\epsilon} $, the closed string 
amplitude can be reinterpreted as 
the open string annulus amplitude. 
This transformation yields 
\begin{eqnarray}
f_1(q) = \frac{1}{2\sqrt{\epsilon}} f_1(\hat q), 
\qquad f_{2,3,4}(q)= f_{4,3,2}(\hat q).
\end{eqnarray}
Besides, we have 
$\Theta_3\left(0 \Big\lvert i\Upsilon \equiv 
\frac{2i\alpha'\epsilon}{\pi R_{\bar i}^2}\right) 
= f_1\left(e^{-\Upsilon}\right) 
f_3{\left(e^{-\Upsilon}\right)}$. 
By expanding the Dedekind and 
$\Theta_3$-functions and also the 
determinant terms for small $\hat q$,
we receive 
\begin{eqnarray}
{\rm Tr} \rho_L^n &\approx& 2^{1-n}
\left(- \frac{\tilde \kappa' V_a T_p^2 
|\det \mathcal M|}{8 \alpha'^{(d_{\ubar{i}}-1)}}\right)^{1-n} 
\left(\frac{4n\epsilon}{(4\epsilon)^n} \right)^{d_{\ubar{i}} 
- (p-6)/2} \exp\left[- \frac{p-6}{48} \left(\frac{1}{n} 
- n\right)\right]
\nonumber\\
&\times& \frac{\mathbf{C}_{(n)}}
{\mathbf{C}^n} \sqrt{\frac{1}{n} 
\left(\frac{\pi R^2_{\bar{i}} }
{2\alpha' \epsilon}\right)^{1-n}} 
\exp\Bigg\{2 n \epsilon \left(\bar q^2_{(n)} 
- \bar q^2\right)\Big[p-6 + {\rm Tr}
(\mathcal Q^T \mathcal Q) \Big]\Bigg\}
\nonumber\\
&\times& \sum_{k_1+k_2+ k_3 \geq 0} (-1)^{k_1} 
\frac{ (n)_{\sum_{i} k_i}}{16^{(k_1+k_2)}k_1! k_2! k_3!}\  
e^{\mathbf{k}\cdot \mathbf{D}} \left[\frac{1}{16}
\Big( e^{\mathbf{D}_{1(n)}} 
- e^{\mathbf{D}_{2 (n)}}\Big) 
- e^{\mathbf{D}_{3 (n)}}+1\right],\qquad 
\label{4.2}
\end{eqnarray}
up to the order $O(\bar q^{4k})=O(e^{-1/\epsilon})$. 
The factor $2^{1-n}$ comes from 
the sum over the spin structures. 
In addition, we defined
\begin{eqnarray}
\mathbf{C} &=& \left(1 + \sum_c \left\{ 
\exp\left[-\frac{1 }{8\epsilon 
\alpha'} \mathcal{L}_c^{\bar{a}} 
\mathcal{L}_c^{\bar{b}} (\delta_{\bar{a} \bar{b}} 
+ \mathcal{F}^{a}_{\ \bar{a}} 
\mathcal{F}_{ \ a \bar{b}})\right] 
\right\} \right)  \prod_{\bar{i}}
\Theta_3 \left( 0 \ \Big\lvert \ 
\frac{i \pi R^2_{\bar{i}}}{2\alpha'\epsilon } \right),\\
\mathbf{D}_1 &=& - \frac{1}{2}(p-6)
\left(\frac{1}{24 \epsilon} 
+ \epsilon \bar q^2\right)  
+ 2 \epsilon \bar q \Big[p-6 
- {\rm Tr} (\mathcal Q^T \mathcal Q)\Big] 
- \ln (- \tilde \kappa'),\\
\mathbf{D}_2 &=& -\frac{5 (p-6)}{96\epsilon} + 2\epsilon 
\bar q\Big[(p-6)\bar q - {\rm Tr}
(\mathcal Q^T \mathcal Q) \Big] 
- \ln (- \tilde \kappa'),\\
\mathbf{D}_3 &=& \ln \left(
- \frac{\tilde \kappa}{\tilde \kappa'}\right) 
- (p-6) \left( \frac{1}{96\epsilon} 
+ 2 \epsilon \bar q\right) 
+ \epsilon \bar q^2 \Big[ \frac{p-6}{2} 
- 2 {\rm Tr}(\mathcal Q^T \mathcal Q)\Big].
\label{4.6}
\end{eqnarray}
In Eq. \eqref{4.2}, we have $\{\bar q_{(n)},\mathbf{C}_{(n)}, 
\mathbf{D}_{\ell (n)}|_{\ell\in\{1,2,3\}}\} 
= \{\bar q, {\mathbf{C}}, 
\mathbf{D}_{\ell}|_{\ell\in\{1,2,3\}}\}
|_{\epsilon \rightarrow n\epsilon}\}$. 
The notation $(n)_{\sum_i k_i}$ represents 
the Pochhammer symbol, and is defined as  
$(n)_x = \Gamma (x+n)/\Gamma(n)$, 
which follows from the multinomial expansion. 

We now compute the LREE. 
This is accomplished by taking the 
limit $n\rightarrow 1$ of the R\'{e}nyi entropy, 
which leads to 
\begin{eqnarray}
\mathbf{S}_{\rm LREE} &\approx& \ln 2 
+ \ln\left(- \frac{\tilde \kappa' 
V_a T_p^2 |\det \mathcal M| \mathbf{C}}
{8 \alpha'^{(d_{\ubar{i}}-1)}}\right) 
\nonumber\\
&+& \left(d_{\ubar{i}} -\frac{p-6}{2}
\right)\left(2\ln2 + \ln \epsilon -1 \right) 
- \frac{p-6}{24 \epsilon}
\nonumber\\
&-&\bar q^2 [p-6 + {\rm Tr} (\mathcal 
Q^T \mathcal Q)] + \frac{1}{2} \ln 
\left(\frac{\pi R_{\bar{i}}^2}{2\alpha' 
\epsilon} \right) 
\nonumber\\
&-& \frac{1}{2\alpha' \epsilon \mathbf{C}}\left\{ 
\frac{1}{4}\sum_{c} \mathcal{L}_c^{\bar{a}} 
\mathcal{L}_c^{\bar{b}} (\delta_{\bar{a} 
\bar{b}} + \mathcal{F}^{a}_{\ \bar{a}} 
\mathcal{F}_{ \ a \bar{b}}) 
+ \left. \sum_{\bar{i}}\frac{i\pi 
R_{\bar{i}}^2}{2\alpha' \epsilon} \right\}
\nonumber\right.\\
&+& \sum_{k_1 + k_2 +k_3 \geq 0} (-1)^{k_1} 
\frac{ (n)_{\sum_{i} k_i}}{16^{(k_1+k_2)}k_1! k_2! k_3!}\  
e^{\mathbf{k}\cdot\mathbf{D}} 
\nonumber\\ 
&\times &\left\{\frac{p-6}{8}
\Bigg[- \frac{1}{12 \epsilon} \left(2 
e^{\mathbf{D}_1} + 5 e^{\mathbf{D}_2} 
+ 16 e^{\mathbf{D}_3}\right) + \epsilon 
\bar q\left(\frac{e^{\mathbf{D}_1}\bar q}{4} 
+ 16 e^{\mathbf{D_3}}\right)\right.
\nonumber\\
&-&\left. \epsilon\bar q \left[ e^{\mathbf{D}1} 
- \bar q \left(e^{\mathbf{D}_2} 
- 8 e^{\mathbf{D}_3}]\right)\right]\Bigg] 
+ \frac{\epsilon \bar q}{8} 
{\rm Tr}\left(\mathcal Q^T \mathcal Q\right) 
\left(e^{\mathbf{D}_1} - e^{\mathbf{D}_2} 
- 16 e^{\mathbf{D}_3}\right) \right\},
\end{eqnarray}
up to the order $O(\bar q^{4k})=O(e^{-1/\epsilon})$,
as in Eq. \eqref{4.2}. 
The first term arises from the summation over 
the spin structures. The second term 
belongs to the boundary entropy, 
associated with the brane. The terms that involve 
$R_{\bar{i}}$ obviously originate from the 
compactification of the spacetime. 
The remaining contributions 
are from the oscillators and the conformal ghosts. 
Although the background and internal 
fields nearly permeate to all terms,  
the dynamics of the brane do not contribute 
to the compactification terms. As discussed in Sec. 
\ref{300}, the quantity $\tilde{\kappa}'$ possesses a 
nonzero value only in specific 
configurations, such as in our case.
As we see, this factor was appeared in
the LREE of our setup. 
 
It should be mentioned that the 
divergence of terms which are proportional to 
$1/\epsilon$ can be justified  
by the sum of all oscillating modes. This is 
more evident when one investigates 
this divergence in the compactification terms. 
This is a result of the fact that 
open strings have higher masses in the  
compact spacetime, and is due to the 
contributions of the quantized momentum. 
It may be neglected in the 
leading terms, which are obtained by 
the lightest open string states.

The LREE in the noncompact spacetime is given by 
\begin{eqnarray}
\mathbf{S}^{\rm non-compact}_{\rm LREE} &\approx& \ln 2 
+ \ln\left(- \frac{\tilde \kappa' V_a T_p^2 
|\det \mathcal M|}{8 \alpha'^{(d_{\ubar{i}}-1)}}\right) 
+ \left(d_{\ubar{i}} -\frac{p-6}{2}\right)\left(2\ln2 
+ \ln \epsilon -1 \right) 
\nonumber\\
&-&\frac{p-6}{24 \epsilon}
-\bar q^2 \left[p-6 + {\rm Tr} (\mathcal Q^T 
\mathcal Q)\right] 
\nonumber\\
&+& \sum_{k_1 + k_2 + k_3 \geq 0} (-1)^{k_1} 
\frac{ (n)_{\sum_{i} k_i}}{14^{(k_1+k_2)}k_1! k_2! k_3!}\  
e^{\mathbf{k}\cdot \mathbf{D}} 
\nonumber\\ 
&\times &\left\{\frac{p-6}{8}
\Bigg[- \frac{1}{12 \epsilon} \left(2 
e^{\mathbf{D}_1} + 5 e^{\mathbf{D}_2} 
+ 16 e^{\mathbf{D}_3}\right) + \epsilon 
\bar q\left(\frac{e^{\mathbf{D}_1}\bar q}{4} 
+ 16 e^{\mathbf{D_3}}\right)\right.
\nonumber\\
&-&\left. \epsilon\bar q\left[ e^{\mathbf{D}_1} 
- \bar q \left(e^{\mathbf{D}_2} 
- 8 e^{\mathbf{D}_3}\right)\right]\Bigg] 
+ \frac{\epsilon \bar q}{8} {\rm Tr}(\mathcal Q^T \mathcal Q) 
\left(e^{\mathbf{D}_1} - e^{\mathbf{D}_2} 
- 16 e^{\mathbf{D}_3}\right) \right\}. \qquad 
\end{eqnarray}

\subsection{The case $p=6$}
\label{410}

Based on Eqs. \eqref{4.2}-\eqref{4.6}, 
the critical brane dimension that 
controls the convergence or divergence 
of the exponential factors is 
$p=6$. Notably, D$6$-branes are 
essential due to their role in generating 
supersymmetric gauge theories \cite{25}, 
contributing to the compactifications of the 
Calabi-Yau spaces \cite{26}, and facilitating 
the non-commutative geometry \cite{27}. 
Additionally, they are associated with the Kaluza-Klein 
monopoles in the M-theory \cite{28}, providing 
a framework for exploring non-perturbative 
effects and topological transitions, crucial 
to the homological mirror symmetry \cite{29}.

The corresponding LREE of a dressed-moving D$6$-brane in 
$\mathbf{T}^n \otimes \mathbb R^{1,9-n}$ finds the feature
\begin{eqnarray}
\mathbf{S}^{(p=6)}_{\rm LREE} &\approx& \ln 2 + 
\ln\left(- \frac{\tilde \kappa' V_6 T_6^2 |\det \mathcal M| 
\mathbf{C}}{8 \alpha'^{(d_{\ubar{i}}-1)}}\right) + d_{\ubar{i}}
\left(2\ln2 + \ln \epsilon -1 \right) - \bar q^2 {\rm Tr} 
(\mathcal Q^T \mathcal Q)
\nonumber\\
&+& \frac{1}{2} \ln \left(\frac{\pi 
R_{\bar{i}}^2}{2\alpha' \epsilon} \right) 
- \frac{1}{2\alpha' \epsilon \mathbf{C}}\left\{ 
\frac{1}{4}\sum_{c} \mathcal{L}_c^{\bar{a}} 
\mathcal{L}_c^{\bar{b}} 
(\delta_{\bar{a} \bar{b}} + \mathcal{F}^{a}_{\ \bar{a}} 
\mathcal{F}_{ \ a \bar{b}}) 
\nonumber\right.\\
&+& \left.  \sum_{\bar{i}}\frac{i\pi 
R_{\bar{i}}^2}{2\alpha' \epsilon} \right\} 
+ \left(\sum_{k_1+ k_2+ k_3 \geq 0} (-1)^{k_1} 
\frac{ (n)_{\sum_{i} k_i}}{16^{(k_1+k_2)}
k_1! k_2! k_3!}\ e^{\mathbf{k}\cdot\mathbf{D}} \right)
\nonumber\\ 
&\times & \frac{\epsilon \bar q}{8} 
{\rm Tr}(\mathcal Q^T \mathcal Q) \left(e^{\mathbf{D}_1} 
- e^{\mathbf{D}_2} - 16 e^{\mathbf{D}_3} \right),
\end{eqnarray}
where for $p=6$ we have
\begin{eqnarray}
&~& \mathbf{D}_1 = \mathbf{D}_2 
= -\left[ 2\epsilon \bar q {\rm Tr} (\mathcal Q^T \mathcal Q) 
+\ln (- \tilde \kappa')\right]\Big |_{p=6},\\
&~& \mathbf{D}_3 = \left[\ln \left(- \frac{\tilde 
\kappa}{\tilde \kappa'}\right) 
-2\epsilon \bar q  {\rm Tr} (\mathcal Q^T \mathcal Q)
\right]\Big |_{p=6}.
\end{eqnarray}

\subsection{A note on the equity  
$\mathbf{S}_{\rm LREE} = \mathbf{S}_{\rm Thermal}$}
\label{420} 

According to the uncertainties that surround 
the definition of the entanglement entropy, 
it is reasonable to explore further 
criteria to select a specific prescription. 
Thus, we focus on the thermodynamical 
entropy. Precisely, we can associate 
the LREE of the brane in our configuration 
with a thermodynamical entropy. 
This analogy is achieved by introducing 
the temperature $T \propto 1/\epsilon$. 
In this prescription, the limit $\epsilon \rightarrow 0$ 
obviously corresponds to the high-temperature 
limit of the thermal system. 

We observed that Eq. \eqref{4.2} completely aligns 
with the thermodynamical entropy, derived 
from the partition function \eqref{11}, in the limit 
$\beta = k_B /T \equiv 2\epsilon \rightarrow 0$. 
This is true even in the presence of the R-R 
sector and the spacetime compactification. 
Despite this satisfactory correspondence, 
it is important to note that these two entropies 
manifestly represent distinct physical 
quantities. Nevertheless, this appealing 
connection may reveal a 
deeper relation between the entanglement 
entropy and its thermodynamical counterpart. 
Additional studies, such as those 
presented in Refs. \cite{30}-\cite{34}, 
have also demonstrated analogous links, 
between the entanglement entropy and the 
first law of thermodynamics.

\section{Conclusions}
\label{500}

In the framework of type IIA/IIB superstring theories, 
we employed the boundary state 
formalism to investigate the LREE of a 
BPS-stable D$p$-brane. The D$p$-brane 
was assumed to move along one of its perpendicular 
directions. Besides, it was  
dressed with the antisymmetric $B$-field 
as well as a $U(1)$ gauge potential. 
The analysis was performed in the
partially compact spacetime.  
To achieve the LREE, the interaction amplitude
between two identical D$p$-branes was computed. 

The LREE obtained a generalized 
form via the presence of various 
parameters in the setup. Consequently, these parameters 
enabled the LREE to be adjustable to any desirable value. 
In addition, we observed that the compactification terms
in the LREE are not influenced 
by the velocity of the brane. 
However, the background and internal fields are present 
in nearly all terms of it. 
We determined that the critical dimension 
of the brane, influencing the 
convergence/divergence of the 
exponential factors, is $p=6$. 
In this special case, the LREE was drastically simplified, 
i.e. all divergences were confined to 
the compactified terms. For a dressed-moving 
D$6$-brane in the non-compact spacetime, 
the LREE can be conveniently computed, which 
reveals the absence of the divergence terms.

Finally, we introduced a temperature 
parameter to the system, which enabled us
to derive the thermodynamical entropy 
through the partition function. 
Remarkably, the thermal entropy is precisely equivalent
to the LREE of the configuration. Similar equivalences 
have also been shown in Refs. \cite{5}, 
\cite{22}, \cite{23}, \cite{30}-\cite{34}. 


\appendix 
\section{The boundary state computations} 
\label{app}

In this appendix, we provide a comprehensive 
review of the boundary state computations. 
The boundary state corresponding to a D$p$-brane with 
the background and internal 
fields is constructed by the well-known 
closed string $\sigma$-model action 
with a single boundary term
\begin{equation}
\label{a1}
S_\sigma = -\dfrac{1}{4 \pi \alpha^\prime} \int_\Sigma
{\rm d}^2 \sigma \left(\sqrt{-h} h^{AB} G_{\mu\nu}
+ \epsilon^{AB} B_{\mu\nu} \right)
\partial_A X^\mu \partial_B X^\nu +
\dfrac{1}{2\pi \alpha^\prime} \int_{\partial\Sigma}
{\rm d}\sigma A_\alpha \partial_\sigma X^\alpha.
\end{equation}
Here, $\mu, \nu \in \{0, 1,\cdots, 9\}$ 
denote the spacetime indices, 
while ``$\alpha$'' and ``$\beta$'' refer to the 
worldvolume directions of the brane. 
We assume a flat spacetime, characterized by the metric 
$\eta_{\mu\nu} = \text{diag}(-1, 1, \cdots, 1)$, 
along with a flat worldsheet metric $h_{AB} = \eta_{AB}$, 
where $A, B \in \{\tau, \sigma\}$. 
In addition, we consider a constant Kalb-Ramond 
field $B_{\mu\nu}$. The Landau gauge is employed, in which the 
gauge potential is specified as 
$A_\alpha = -\frac{1}{2} F_{\alpha\beta} 
X^\beta$, where $F_{\alpha\beta}$ 
represents the constant field strength.
Note that $\{\alpha\}=\{0\}\bigcup\{a\}$, and 
$a\in \{1, 2, \cdots , p\}$.

The following boundary state equations, as well as
the equation of motion, are conveniently 
obtained by setting the variation of the action to zero
\begin{eqnarray}
&&\Big(\partial_\tau X^\alpha 
+\mathcal{F}^\alpha_{\ \ \beta}
\partial_\sigma X^\beta \Big)_{\tau=0} | B_x\rangle = 0 ,
\label{a2}\\
&&\left(X^i - y^i \right)_{\tau=0} | B_x\rangle = 0.
\label{a3}
\end{eqnarray}

To impose a transverse velocity $\mathcal V$ on the D$p$-brane, 
one must apply the Lorentz boost transformations 
to the boundary state equations. 
Let $x^{i_{\mathcal V}}$ denote the 
transverse coordinate along the boost 
direction. Consequently, the boundary state equations
of the boosted brane take the features  
\begin{eqnarray}
&&\Big[\partial_\tau (X^0 - \mathcal V X^{i_{\mathcal V}}) 
+ \mathcal{F}^0_{\ \ a } \partial_\sigma 
X^a \Big]_{\tau=0} |B_x\rangle = 0, 
\nonumber \\
&&\Big(\partial_\tau X^{a} 
+ \gamma^2 \mathcal{F}^{a}_{\ \ 0} 
\partial_\sigma (X^0 - \mathcal V X^{i_{\mathcal V}}) 
+ \mathcal{F}^{a}_{\ \ b} 
\partial_\sigma X^{b} \Big)_{\tau=0} 
|B_x\rangle = 0, 
\nonumber \\
&&\Big(X^{i_{\mathcal V}} - \mathcal V X^0 
- y^{i_{\mathcal V}} \Big)_{\tau=0} 
|B_x\rangle = 0, 
\nonumber \\
&&\left(X^i - y^i \right)_{\tau=0} 
|B_x\rangle = 0, \quad i \ne i_{\mathcal V}, 
\label{a4}
\end{eqnarray}
where $\gamma = (1-\mathcal V^2)^{-1/2}$ 
represents the boosting factor.

The general solution of the equation of 
motion for the closed string is
\begin{eqnarray}
X^\mu (\sigma, \tau) = x^\mu + 2\alpha' 
p^\mu \tau + 2 \mathcal 
L^\mu \sigma + i \sqrt{\alpha'/2} 
\sum_{m\ne 0}m^{-1}\Big(\alpha^{\mu}_m 
e^{-2im (\tau -\sigma)} 
+ \tilde\alpha^\mu_m e^{-2im (\tau +\sigma)}\Big).\ \ 
\label{a5}
\end{eqnarray}
For the non-compact directions, $\mathcal L^\mu$ 
obviously vanishes. 
However, for the compactified directions, 
$\mathcal L^\mu$ is given by 
$\mathcal L^\mu = (NR)^\mu$, and the corresponding momentum is 
$p^\mu = (M/R)^\mu$. Here, $N^\mu$, 
$M^\mu$ and $R^\mu$ represent the 
winding number, momentum number of the 
closed string state and radius of the 
compactification for $X^\mu$ direction, respectively. 

By substituting Eq. \eqref{a5} into the 
boosted boundary state equations \eqref{a4}, 
one acquires these equations in terms of the 
oscillators 
\begin{eqnarray}
&&\Big(\alpha^0_m - \mathcal V \alpha^{i_{\mathcal V}}_m -
\mathcal{F}^0_{\ \ a} \alpha^{a}_m
+ \tilde{\alpha}^0_{-m} - \mathcal V
\tilde{\alpha}^{i_{\mathcal V}}_{-m}
+ \mathcal{F}^0_{\ \ a}
\tilde{\alpha}^{a}_{-m}\Big)
|B_x\rangle_{\rm osc}= 0,
\nonumber \\
&&\Big\{\alpha^{a}_m - \gamma^2
\mathcal{F}^{a}_{\ \ 0} \Big[\alpha^0_m
- \mathcal V (\alpha^{i_{\mathcal V}}_m 
-\tilde{\alpha}^{i_{\mathcal V}}_{-m})
- \tilde{\alpha}^0_{-m}\Big] +\tilde{\alpha}^{a}_{-m}
-\mathcal{F}^{a}_{\ \ b}
(\alpha^b_{m}- \tilde{\alpha}^b_{-m})\Big\}
|B_x\rangle_{\rm osc} = 0,
\nonumber\\
&& \left[\alpha^{i_{\mathcal V}}_{m} 
- \tilde{\alpha}^{i_{\mathcal V}}_{-m}
-\mathcal V  (\alpha^0_m - \tilde{\alpha}^0_{-m})\right]
|B_x\rangle_{\rm osc} = 0,
\nonumber \\
&& \left(\alpha^i_{m} - \tilde{\alpha}^i_{-m}\right)
|B_x\rangle_{\rm osc} = 0 , \qquad i\ne i_{\mathcal V}.
\label{a6}
\end{eqnarray}
These equations conveniently can be rewritten 
in the collective form 
$\Big(\alpha_m^\mu + \mathcal O^\mu_{\ \nu} 
\tilde\alpha_{-m}^\nu\Big)|B_x\rangle_{\rm osc} = 0 $. 
Employing the coherent state formalism, the solution for the 
oscillatory portion of the boundary 
state $|B_x\rangle_{\rm osc}$ is given by
\begin{eqnarray}
|B_x\rangle_{\rm osc} &=& \sqrt{- \det \mathcal M}
\exp \left[ - \sum_{m=1}^{\infty} \left(\dfrac{1}{m}
\alpha^\mu_{-m} \mathcal O_{\mu \nu}
\tilde \alpha^{\nu}_{-m}\right)\right]
{|0_\alpha, 0_{\tilde \alpha}\rangle}\;,
\label{a7}
\end{eqnarray}
where the prefactor $\sqrt{-\det \mathcal{M}}$ arises from the 
disk partition function \cite{20} 
as a result of a regularization scheme. 
The matrix $\mathcal O_{\mu\nu}$ possesses 
the following definition
\begin{eqnarray}
\mathcal O_{\mu\nu} &\equiv & 
\Big(\mathcal Q_{\lambda\lambda'}
\equiv(\mathcal M^{-1}\mathcal N)_{\lambda\lambda'}
|_{\lambda,\lambda' \in \{\alpha,i_\mathcal V\}}\;
,\;-\delta_{ij}|_{i,j \ne i_\mathcal V}  \Big),
\label{2.6}
\end{eqnarray}
in which the matrices $\mathcal M$ and $\mathcal N$ 
are given by
\begin{eqnarray}
\mathcal M^0_{\ \lambda} = \left(\delta^0_{\ \lambda} -
\mathcal V \delta^{i_{\mathcal V}}_{\ \lambda} 
- \mathcal{F}^0_{\ a}
\delta^{a}_{\ \lambda}\right), &\quad& 
\mathcal N^0_{\ \lambda} 
= \left(\delta^0_{\ \lambda}
- \mathcal V \delta^{i_{\mathcal V}}_{\ \lambda}
+ \mathcal{F}^0_{\ a}
\delta^{a}_{\ \lambda}\right),
\nonumber\\
\mathcal M^{a}_{\ \lambda} =
\delta^{a}_{\ \lambda}
-\gamma^2 \mathcal{F}^{a}_{\ 0}(\delta^{0}_{\ \lambda}
-\mathcal V\delta^{i_{\mathcal V}}_{\ \lambda})
-\mathcal{F}^{a}_{\;\;\;b}
\delta^{b}_{\ \lambda}, &\quad& \mathcal N^{a}_{\ \lambda} =
\delta^{a}_{\ \lambda}
+ \gamma^2 \mathcal{F}^{a}_{\ 0}
(\delta^{0}_{\ \lambda}
- \mathcal V \delta^{i_{\mathcal V}}_{\ \lambda})
+ \mathcal{F}^{a}_{\;\;\;b}
\delta^{b}_{\ \lambda},
\nonumber\\
\mathcal M^{i_{\mathcal V}}_{\ \lambda} 
= \delta^{i_{\mathcal V}}_{\ \lambda}
- \mathcal V \delta^{i_{\mathcal V}}_{\ \lambda}, 
&\quad& \mathcal 
N^{i_{\mathcal V}}_{\ \lambda} = 
-\delta^{i_{\mathcal V}}_{\ \lambda}
+ \mathcal V \delta^{i_{\mathcal V}}_{\ \lambda}.
\label{a9} 
\end{eqnarray}

Eq. \eqref{a5} implies that the 
effect of the compactification is 
exclusively manifested in the zero-mode 
sector of the boundary state. 
The zero-modes contribution to the boundary state equations 
are given by
\begin{eqnarray}
&&\left(\alpha'p^0 -\alpha' \mathcal V 
p^{i_{\mathcal V}} + \mathcal 
F^0_{\ a}\mathcal L^a\right)
|B_x\rangle_{0} = 0,
\nonumber \\
&&\left(\alpha' p^a + \mathcal F^a_{\ b} 
\mathcal{L}^b\right)|B_x\rangle_{0}= 0,
\nonumber \\
&&\left(x^{i_{\mathcal V}} - \mathcal V x^0 
- y^{i_{\mathcal V}}\right)
|B_x\rangle_{0} = 0,
\nonumber \\
&&\left(x^i - y^i\right) |B_x\rangle_{0} = 0 ,
\qquad i\ne i_{\mathcal V}.
\label{a10}
\end{eqnarray}
By employing the formal quantum mechanical methods, 
the zero-mode portion of the boundary state takes the solution
\begin{eqnarray}
|B_x\rangle_{0} &=& \delta(x^{i_{\mathcal V}} 
- \mathcal V x^0 
- y^{i_{\mathcal V}}) \ket{p^{i_{\mathcal V}}_L 
= p^{i_{\mathcal V}}_R 
= \frac{1}{2}\mathcal V p^0}
\nonumber\\
&\times &\left(\prod_{i \ne i_{\mathcal V}}\delta(x^i-y^i)  
\ket{p^i_L=p^i_R=0}\right) \left(\prod_{a=1}^p 
\ket{p^a_L= p^a_R = 0}\right) 
\label{a11}.
\end{eqnarray}
According to Eqs. \eqref{a10} and \eqref{a11}, 
the momentum components find the nonzero quantized values
\begin{eqnarray}
p^a &=& \alpha'^{-1} \mathcal 
F^a_{\ \bar b}\mathcal L^{\bar b}, \\
p^0 &=& -\gamma^2 \alpha'^{-1} 
\mathcal F^0_{\ \bar a}\mathcal L^{\bar a},\\
p^{i_{\mathcal V}} &=& -\gamma^2 
\mathcal V\alpha'^{-1} \mathcal 
F^0_{\ \bar a}\mathcal L^{\bar a}.
\end{eqnarray}
These relations prominently reveal that one can always 
apply summation on the winding numbers  
instead of summation on the momentum numbers. 
This is particularly useful when one 
computes the interaction amplitude.

The supersymmetric extension of the action \eqref{a1}, 
under the global worldsheet supersymmetry,
induces the following transformations
to the bosonic boundary conditions 
\begin{eqnarray}
\partial_+ X^\mu (\sigma, \tau) 
&\mapsto & - i \eta \psi^\mu_+(\sigma, \tau),
\nonumber\\
\partial_- X^\mu (\sigma, \tau) 
&\mapsto &  \psi^\mu_-(\sigma, \tau).
\label{a15}
\end{eqnarray}
Note that $\eta = \pm 1$ denotes the 
GSO projection parameter, and 
$\partial_\pm \equiv \frac{1}{2}
(\partial_\tau \pm \partial_\sigma)$.

Let $d_m^\mu$ ($b_r^\mu$) represent 
the fermionic oscillators in the R-R 
(NS-NS) sector. Hence, we receive 
\begin{eqnarray}
|B_\psi, \eta\rangle_{\rm NS} 
&=& \exp \left(i\eta \sum_{r\geq 1/2} 
b^\mu_{-r} \mathcal O_{\mu\nu}\tilde 
b_{-r}^\nu \right)|0\rangle_{\rm NS},
\label{a16}\\
|B_\psi, \eta\rangle_{\rm R} 
&=&\frac{1}{\sqrt{-\det \mathcal M}} \exp 
\left(i\eta \sum_{n=1}^\infty d^\mu_{-n} 
\mathcal O_{\mu\nu}\tilde 
d_{-n}^\nu \right)|B^0_\psi, \eta\rangle_{\rm R},	
\label{a17}
\end{eqnarray}
where ``r'' belongs to the positive half integers.
In contrast to the bosonic case \eqref{a7}, 
the Grassmannian nature of the 
fermionic oscillators imposed the inverse of the determinant.
The explicit expression for the zero-mode boundary state 
$ |B_\psi^{0}, \eta \rangle_{\text{R}}$ possesses the feature 
\begin{equation}
|B_\psi^{0}, \eta \rangle_{\text{R}} 
= \gamma \left[ C\left(\Gamma^{0} 
+ {\mathcal V} \Gamma^{i_{\mathcal V}}\right) 
\Gamma^1 \dots \Gamma^p 
\frac{1 + i\eta \Gamma^{11}}{1 + i\eta} 
\mathbb{G} \right]_{AB} 
|\mathcal S_A\rangle \otimes |\tilde{\mathcal S}_B\rangle,
\label{a18}
\end{equation}
where ``$C$'' represents the charge 
conjugation matrix, associated 
with the group $SO(1, 9)$, and $ |\mathcal S_A\rangle $ and 
$ |\tilde{\mathcal S}_B\rangle $ 
are the corresponding spinor states of this group. 
The matrix $\mathbb{G}_{32\times 32}$ 
satisfies the following equation 
\begin{equation}
\Gamma^\lambda \mathbb G 
- \mathcal Q^\lambda_{\;\;\lambda'} 
\mathbb G \Gamma^{\lambda'} 
- \mathcal V \Gamma^{i_{\mathcal V}} 
\Gamma^\lambda \Gamma^{0} \mathbb G 
- \mathcal V \Gamma^{i_{\mathcal V}} 
\Gamma^{0} \mathcal Q^\lambda_{\;\;\lambda'} 
\mathbb G \Gamma^{\lambda'} = 0. 
\label{a19}
\end{equation}
The algebra of the Dirac matrices 
enables us to conveniently recast this equation in 
the following appropriate form
\begin{equation}
\Gamma^\lambda(\mathbb I + \mathcal V 
\Gamma^{i_{\mathcal V}} \Gamma^{0} ) 
\mathbb G - \mathcal Q^\lambda_{\;\;\lambda'} 
(\mathbb I + \mathcal V 
\Gamma^{i_{\mathcal V}} \Gamma^{0} )
\mathbb G \Gamma^{\lambda'} 
- 2 \mathcal V \eta^{i_{\mathcal V}\lambda} 
\Gamma^{0} \mathbb G=0.
\end{equation}
Consequently, the solution for $\mathbb G$ 
is explicitly given by
\begin{eqnarray}
\mathbb G &=& \dfrac{*\exp\left(2^{-1} 
\hat\Phi_{\lambda\lambda'} 
\Gamma^\lambda \Gamma^{\lambda'}\right)*}{{\mathbb I} 
+ \mathcal V \Gamma^{i_{\mathcal V}} \Gamma^{0} 
- 2 \mathcal V \Gamma^{i_{\mathcal V}} \Gamma^{0} 
\left[{\mathbb I} 
+ (\Delta \mathcal Q)^{i_{\mathcal V}}_{\;\;\lambda'} 
\Gamma^{i_{\mathcal V}} \Gamma^{\lambda'}\right]^{-1}},
\label{a21}
\end{eqnarray}
where $\hat\Phi\equiv 2^{-1} (\Phi - \Phi^{\rm T})$, $
\Phi_{\lambda\lambda'} \equiv 
\left[(\Delta \mathcal Q+{\mathbb I})^{-1}
(\Delta \mathcal Q-{\mathbb I})\right]_{\lambda\lambda'} $, 
and the matrix $\Delta$ is defined as 
$ \Delta^{\lambda}_{\;\;\lambda'} 
= (\delta^{\alpha}_{\;\;\beta}, 
-1|_{\lambda,\lambda'=i_{\mathcal V}})$ 
with $\Delta^{\alpha}_{\;\;i_\mathcal{V}} = 0 $. 
The notation $*~~* $ indicates that the exponential  
should be expanded such that all 
the Dirac matrices anticommute. 
Consequently, the expansion obviously  
contains a finite number of terms. 

The total boundary state in each sector is given by
\begin{eqnarray}
|B , \eta \rangle^{\rm tot}_{\rm NS(R)} =
\frac{T_p}{2} |B_x \rangle
|B_{\rm gh} \rangle |B_\psi , \eta \rangle_{\rm NS(R)}
|B , \eta \rangle_{\rm sgh},
\end{eqnarray}
where $T_p$ is the tension of the D$p$-brane, and 
``gh'' and ``sgh'' indicate the conformal and 
super-conformal ghosts, respectively. 


\end{document}